# Combining Representation Learning with Tensor Factorization for Risk Factor Analysis -- an application to Epilepsy and Alzheimer's disease

Yejin Kim[1], Samden Lhatoo[2], Guo-Qiang Zhang[1,2], Luyao Chen[1], Xiaoqian Jiang[1]
[1]School of Biomedical Informatics, University of Texas Health Science Center at Houston
[2]McGovern School of Medicine, University of Texas Health Science Center at Houston

## ABSTRACT

Existing studies consider Alzheimer's disease (AD) a comorbidity of epilepsy, but also recognize epilepsy to occur more frequently in patients with AD than those without. The goal of this paper is to understand the relationship between epilepsy and AD by studying causal relations among subgroups of epilepsy patients. We develop an approach combining representation learning with tensor factorization to provide an in-depth analysis of the risk factors among epilepsy patients for AD. An epilepsy-AD cohort of ~600,000 patients were extracted from Cerner Health Facts data (50M patients). Our experimental results not only suggested a causal relationship between epilepsy and later onset of AD ($p$= 1.92e-51), but also identified five epilepsy subgroups with distinct phenotypic patterns leading to AD. While such findings are preliminary, the proposed method combining representation learning with tensor factorization seems to be an effective approach for risk factor analysis.

## INTRODUCTION

Epilepsy and Alzheimer's disease (AD) are among two of the most prevalent serious neurological disorders [1,2]. While incident of epilepsy is most common in the elderly, it is increasingly recognized that seizures are further over-represented in the population with AD [3,4], occurring in up to 64% of closely monitored patients [5]. Seizures may predate AD diagnosis; epileptiform activity can precede amyloid B plaque deposition in transgenic mouse models of AD. Subtle seizure types such as brief focal unaware seizures may go undiagnosed in patients [6], lowering quality of life and predisposing to morbidity and mortality. There is also concern that seizures may hasten cognitive decline, exaggerate the effects of AD [7] and reduce life expectancy [8]. Neuropathological studies have demonstrated the presence of age related amyloid B plaque in resected temporal lobe tissue of epilepsy patients without dementia, compared to control autopsies without epilepsy [9]. Chronic epilepsy is associated with increased tau neurofibrillary tangles at mid-Braak stages in patients aged 40–65 years [10], suggesting common pathophysiological pathways. The relationship between epilepsy and AD requires further elucidation. Health claims data offer a unique perspective to study such complicated relationships. The key is to make good use of the appropriate information to identify critical discriminative patterns from the big health claims data while overcoming bias, confounder, and imprecise onset time point for diseases.

The goal of this paper is to understand the relationship between epilepsy and AD by studying causal relations among subgroups of epilepsy patients. We propose an approach combining representation learning with tensor factorization in order to provide an in-depth analysis of the risk factors among epilepsy patients for AD.

## RELATED WORK

The most relevant technology to our work is computational phenotyping models. These models identify latent groups of similar features (i.e., diagnosis, medication, and labs) that lie in a similar context, with the aim of grouping patients with similar conditions and providing fine-grained stratification into patient cohorts (subgroups).  Such techniques provide an alternative evidence-based perspective to look at retrospective data, especially in understanding complex cohort. A thread of early publications were named after different gems: Marble [11], Rubik [12], Granite [13]. Using entity co-occurrence information (e.g., comorbidity and co-prescribed medications in the same encounter) as the observation, these methods leverage tensor factorization to decompose high dimensional observational tensor into smaller, lower-ranked ones that are more suitable for interpretation. Each of these methods focuses on a slightly different aspect to consider prior/expert knowledge, data sparsity, and phenotypic diversity in order to identify interesting phenotype groups that are clinically meaningful. Essentially, these are all unsupervised soft clustering models that try to stratify the data in low dimensional manifolds. Because of the lack of common discriminative objectives in these clustering approaches, the final outcome, while statistically meaningful, may not always provide clinical insights. A recent study [14] improved this mechanism by introducing a supervision term and a contextual similarity regularization term, enabling tensor factorization to be performed discriminatively in order to minimize a classification loss function towards outcome of interest (e.g., disease onset, mortality, readmission). However, this method considers co-occurrence as a single source of information to be modeled, neglecting the temporal nature of such information as observations change over time.

On the other hand, graph decomposition based methods for computational phenotyping has been developed [15] to model the disease progression of individual patients as separate graphs. Each node in the graph represents a medical event (diagnosis, prescription, etc.) and each edge indicates a (longitudinal) transition relationship (e.g., delirium is diagnosed in the next visit after the diagnosis of AD). Instead of modeling co-occurrences, graph decomposition approaches seek to identify the most representative subgraphs to explain the disease progression. A set of core subgraphs are extracted so that their linear combination can approximate the graph of individual patients. But unfortunately, this method does not allow the integration of co-occurrence information because the edge of the graph can only represent one kind of relationship (either transition or co-occurrence). In general, there is no one-size-fits-all model as either encoding mechanism might lead to cross-sectional or longitudinal information loss.

The latest work by Yin *et. al.* [16] utilized deep learning to model the dynamic temporal states as hidden factors with recursive neural networks (RNN) while keeping co-occurrence patterns encoded in the tensor for joint optimization. This is the first work to harmonize information of both sources and demonstrates improved performance. One limitation of this model is that it assumes that phenotypes are static combinations of different medical events instead of evolving patterns (i.e., using the phenotype membership change over time to represent the patient's state dynamics). For chronic disease like AD, such an assumption may not be sound as temporal pattern is most important to characterizing patient's state. Therefore, we need to model the transition relationships in a more explicit manner (while still considering the co-occurrence information). To the best of our knowledge, we are the first to utilize representation learning (based on co-occurrences) and tensor factorization (based on transition) in an orthogonal manner to complement each other for explicit temporal phenotyping study.

**DATASET**

We extracted the Epilepsy-AD cohort as a subset of Cerner Health Facts data (2000 ~ 2014) subscribed by UTHealth [17], which has ~50M patients from 600 Cerner client hospitals. The data elements include demographics, encounters, diagnoses, procedures, medication orders, medication administration. The epilepsy patients are identified by ICD-9 code 345.x and the AD patients are identified by 331.0. The experiment section has detailed breakdown of our cohort.

**METHOD**

**Evaluate the causal relationship between Epilepsy and Alzheimer's disease**
Our first step is to verify the causal relationship between Epilepsy and AD using our data. It is also a step to confirm that our data are sufficient to support the downstream phenotyping study. For this purpose, we first selected epilepsy patients with and without developing AD, followed by using propensity score matching (PSM) algorithm [18] to find a comparable cohort of patients without epilepsy (matched on demographics and three major epilepsy risk factors: brain injury, brain tumor, and stroke). In the matching process, we select as many candidates as possible while meeting the boundary condition that each matching variable should not deviate more than 5% after matching. Finally, we conduct $\chi^2$ test to check the significance of the outcome.

**Temporal phenotype mining that integrates co-occurrence and transition patterns**
We propose an innovative framework to capture two important aspect of patient information trajectory (1) *co-occurrence* pattern, (2) *(longitudinal) transition* pattern. As illustrated in Figure 1, the electronic health records (EHR) can be represented as a sequence of set of entities (e.g., medications, diagnoses, lab tests, conditions). Within each visit (encounter), a patient will receive a set of medication and diagnosis, which order does not matter. This is followed by another visit and we might observe changes in medication and diagnosis, which forms the so called transition pattern, providing clues to the progression of the disease or other changes to the patient's health condition.

We will model the co-occurrence (within the same encounter) of entities using a representation learning mechanism and model the transition (between encounters) of entities directly using tensors factorization. A joint optimization framework that considers both the trajectory similarity (in transition) and entity-level similarity (in co-occurrence) is developed to mine critical temporal phenotypes that might explain the difference between epilepsy patients who are more likely to develop AD (AD likely) vs. those who are less likely to develop AD (AD unlikely).

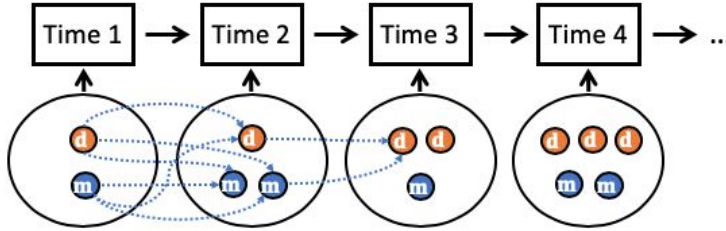

**Figure 1:** A spatial-temporal view of patient's treatment pathway. Entities in the bag of each visit include a set of medications and diagnoses, which are subject to change from time to time. Each edge indicate a tradition pattern.

*Learning co-occurrence information by representation learning*

EHR data are sparse as different patients have different visit frequencies and encounters in their clinical pathway. On the other hand, there are a massive number of entities, which present difficulties for simple statistical machine learning models that rely on frequent pattern mining (including tensor based computational phenotyping). A mitigation approach is to view entities in their context and represent them as dense vectors (e.g., absorbing co-occurrence information of the neighbors) to feed downstream models for analysis. Here we will adopt distributed embedding approach to assemble the information from the neighborhood of individual entities. We will choose the skip-gram approach for efficiency. Assuming that the training data are assembled in the form of visit $t$: $\{m_1, m_2, \cdots, m_K, d_1, d_2, \cdots, d_L\}$ for each encounter of every patient, where $m_k$ and $d_l$ correspond to the medication and diagnosis in this list. The likelihood model for each $m_i$ is:

$$\frac{1}{K+L} \sum p(m_1, m_2, \cdots, m_{i-1}, m_i, \cdots, m_K, d_1, d_2, \cdots, d_L) = \frac{1}{K+L} \sum_{m_k, d_l \in N(m_i)} \log p(m_k|m_i) + \log p(d_l|m_i)$$

essentially predicting everything else $m_k, d_l \in N(m_i)$ within this encounter based on $m_i$. In order to model such probability, we use a neural network to learn within the latent embedding space $v$, for which the dimensionality $d$ is pre-determined by the modeler and each entity $m_i$ is represented by a dense vector $v_{m_i}$. The conditional probability

$$p(m_k|m_i) = \frac{\exp(v_{m_k}^T v_{m_i})}{\sum_m \exp(v_m^T v_{m_i})} \quad (1)$$

is to use all the medications to infer the likelihood of $m_i$ in a softmax fashion. Similarly, we can do it for the $p(d_l|m_i)$ by considering all the diagnoses. The challenge to optimize the above likelihood directly is the big domain of medication and diagnosis (i.e., thousands of entities), making the computation intractable. A practical approach is to use an approximation technique call noise contrastive sampling (NCE) [19], which respects the original distribution of the data. In NCE, we basically get the $L$ false samples for each positive sample. In the case of our Eq. (1), we replace the denominator with $\sum_l \exp(v_l^T v_{m_i})$ where $v_l$ is sampled from entities that are not observed in the same encounter of $m_i$ while still satisfying the general distribution of medication. Solving this large optimization problem with $L+1$ samples for each entity, we will obtain the final embedding vectors.

*Learning transition information by tensor factorization*

In addition to the co-occurrence within encounter, transition from one entity to another entity in consecutive encounters is the key to model the progression of disease. We first derive each individual patient's transition from previous visit to next visit. One patient, P1 follows a trajectory of entities, for example, Visit 1: $\{d_1, m_1\}$ → Visit 2: $\{d_2, m_1\}$ → Visit3: $\{d_2, m_2\}$. We represent this trajectory as transitions between pairs of entities in consecutive visits: $d_1 \to d_2$, $d_1 \to m_1$, $m_1 \to d_2$, $m_1 \to m_2$, and so on. This pairwise transition can be represented as a matrix where each value refers to counts or normalized counts (i.e., transition probability) of the transitional pair across all encounters.

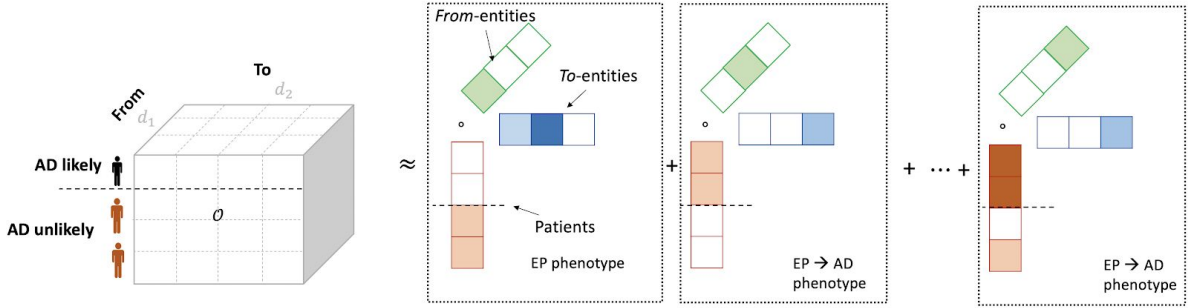

**Figure 2.** Tensor representing transition of entities across all patients. EP=epilepsy phenotype unlikely to develop AD , EP→ AD = epilepsy phenotype likely to develop AD.

However, this individual patient's transition matrix contains only each individual's different progression of diseases. To aggregate the different transitions into a population-level view, we stack the transition matrices into a 3-order matrix, i.e., tensor (Figure 2. left). This 3-order tensor $O$ has a shape of $I \times J \times J$, where $I$ = number of patients, $J$ = number of entities. Then, we decompose the tensor into latent dimensions to derive *phenotypes*, which is defined as underlying transitions from a group (cluster) of relevant entities to another group of relevant entities. The transition between groups is useful to understand generalized progression of diseases (in our case, progression of EP to AD), where the transition from one entity to another does not provide a context in which the transitions happen. The most widely used tensor decomposition is CP method. A 3-order tensor $O$ with a shape of $I \times J \times J$ is rank-one tensor if it is an outer product of three vectors $a, b, c$, i.e., $O = a \circ b \circ c$ where $\circ$ means the vector outer product. $O_{ijk}$, the element at $(i, j, k)$ in the tensor $O$, is computed as a product of elements in the vector, i.e., $O_{ijk} = a_i b_j c_k$. Tensor factorization (TF) is a dimensionality reduction approach that represents the original tensor as latent dimensions. The CP model approximates the original observed tensor $O$ as a linear combination of rank-one tensors [20]; that is, the 3-order tensor $O$ is decomposed as minimizing difference between observed tensor and approximated tensor as

$$L = \| O - \sum_{r=1}^{R} a_r \circ b_r \circ c_r \|^2$$

where a positive integer $R$ is the rank (i.e., number of phenotypes), $a_r$, $b_r$, $c_r$ are $r$-th column vectors in matrix $A$, $B$, $C$ with shape of $I \times R$, $J \times R$, $J \times R$, respectively. Here, $A$, $B$, $C$ are called as factor matrices, and A, B, C corresponds to patient, entity (where the transition comes *from*), entity (where the transition goes *to*). The $R$ latent dimensions in the $R$ columns of factor matrices define phenotypes. That is, a phenotype consists of the *from*-entities in the column of $B$ and the to-entities in the column of $C$. Individual entities are involved to the phenotype with different degree of *membership*, and the amount of membership is stored in the columns of $B$ and $C$. Likewise, individual patients have characteristics of the $R$ phenotypes with different extent of *membership*, and the amount of membership is stored in the row of $A$. On top of the basic TF, we can add several constraints and regularizers to enhance interpretability and discriminative power of phenotypes. Because the tensor $O$ contains non-negative data (counts or transition probability), we set non-negative constraints $A, B, C \geq 0$ for interpretability of latent dimensions. A $l_1$ norm regularizer to the factor matrix enhances interpretability via compact patterns as shrinking the less important coefficients to zero:

$$L + \lambda \cdot (\|A\|_1 + \|B\|_1 + \|C\|_1)$$

where $\lambda$ is a weight parameter to balance the tensor error and the $l_1$ norm loss. A supervised regularizer to the factor matrix $A$ (which defines how much each patient is involved in the phenotypes) boosts discriminative power by separating patients into either EP+AD or EP [14]. The supervised TF adds logistic regression regularizer as

$$L - \mu \cdot \log P(A, y|\theta) = L - \mu \cdot \frac{1}{1+exp(-y \cdot \theta A)}$$

where µ is a weight parameter to balance the tensor error and the loss on regularizer, θ is parameter for discriminative logistic regression objective, and $y$ is label ($y = 1$ if EP+AD; -1 EP only). Therefore, our final objective function is integrating both regularizers: $L - \mu \cdot \log P(A, y|\theta) + \lambda \cdot (\|A\|_1 + \|B\|_1 + \|C\|_1)$.

*Jointly optimizing transition information and co-occurrence similarity by coupled tensor factorization*
We incorporate the co-occurrence representation to the transition pattern to derive temporal phenotype that considers both aspects within- and between- visits. We aim to make phenotypes respect the co-occurrence similarity while reflecting the temporal dynamics of transition pattern. We first derive pairwise similarity between entities using the co-occurrence representation: $cosine(v1, v2)$ where $v1, v2$ is an arbitrary embedding vector. Then, we build similarity matrix $S$ with a shape of $J \times J$. When there are multiple sources of information, coupled matrix-tensor factorization can jointly learn the factor matrices that reflect both sides of information [14]. Because the transition tensor $O$ and the similarity matrix $S$ share the same entity information ($B$ and $C$), the loss function becomes

$$L = \| O - \sum_{r=1}^{R} a_r \circ b_r \circ c_r \|^2 + \gamma(\|S - BB^T\|^2 + \|S - CC^T\|^2) \qquad (2)$$

where $\gamma$ is a weighting constant.

**EXPERIMENT**
We will discuss the results of our experiments in two different sections. The first section is to show that there is a strong causal relationship between epilepsy and AD based on our data. The second section will focus on the epilepsy patients to analyze the computational phenotypes of AD likely and AD unlikely.

**RESULTS**
**Statistical evaluation on the causal relationship between epilepsy and AD**
We use PSM algorithm to avoid data selection bias when building the cohort. We expect the case and control patients have the same age (at least 45 years old) when we start the observation and the same length of observation windows (at least 0.5 years). As AD patients normally start to develop the symptoms from age 65, we only select cases with first epilepsy onset after 55 years of age. Our PSM includes the demographic data of gender, race, and ethnicity. We also made sure the case and control patients have the same epilepsy risk factors - brain injury, brain tumor, and stroke. These factors can produce 61% accuracy when predicting epilepsy with the logistic regression model. We fine-tuned the PSM method to make sure the standardized bias within 5% in the matched data. Table 1 shows the cohort size at different stages of our pre-processing. Figure 3 lists the discrepancy of various properties, showing that the differences are well-controlled under 5% after our matching process while keeping a large number of samples from the case group. After the matching, we use the $\chi^2$ test to evaluate significance as to whether epilepsy patients have a higher chance of subsequently developing Alzheimer's disease. Our result indicates that such a causal relation is significant with a *p*-value =1.92e-51 (Table 2).

**Table 1**: Cohort of epilepsy patients (case) and controls

|  | **Epilepsy Patients (case)** | **Control Patients** |
|---|---:|---:|
| Original data | 59,101 | 600,000 |
| After clean up (delete records with incomplete data) | 58,728 | 598,048 |
| After propensity score matching | 58,432 | 58,432 |
| After removing patients with AD onset before epilepsy | 55,770 | 57,785 |

| Properties | Original data | Matched data |
|---|---|---|
| Age at the start of observation | 41.26% | 2.18% |
| Observation window | 15.72% | 0.51% |
| Gender | 3.74% | 0.16% |
| Race | 4.78% | 1.29% |
| Ethnicity | 1.91% | 1.77% |
| Brain injury | 8.46% | 4.65% |
| Brain tumor | 11.3% | 2.97% |
| Stroke | 22.38% | 0.58% |

(a) Standardized variance before and after the propensity score matching

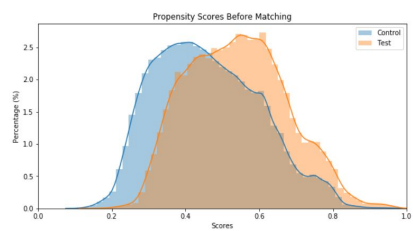

(b) Propensity score before matching

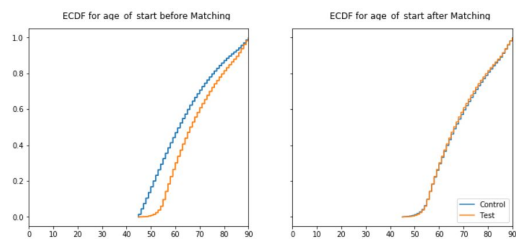

(c) Age distribution before and after matching

**Figure 3:** The distributional information before and after the propensity score matching

**Table 2:** Chi-square test on the causal relationship between Epilepsy and Alzheimer. Chi-squared value was 227.67 with p-value of (1.92e-51) < 0.0001

|  | No Epilepsy | Epilepsy |
|---|---|---|
| No Alzheimer | 57,123 | 54,481 |
| Alzheimer | 662 | 1289 |

**Computational phenotyping results and discussion**

We extracted a subset of data from the case in the previous section to study those epilepsy patients who developed AD and those who did not develop AD in a similar observation window. We select all 1,289 patients who developed AD after epilepsy as cases and randomly selected another 6,500 patients without AD as controls. We extracted 97 diagnosis codes (ICD-9) and 106 drugs, which are associated with more than 5% of these patients (Table 3).

**Table 3:** Statistics of Cohort for phenotyping. Note that this is a subset of the case group for causality analysis

| Features | Cases<br>Developed AD after epilepsy onset | Controls<br>Did not develop AD after epilepsy onset |
|---|---|---|
| Number of Patients | 1,241 | 5,276 |
| Age at the start of observation | 74.87±8.92 | 68.09±10.79 |
| Observation window | 3.88±227 | 3.73±2.29 |
| Gender - Female | 41.7% | 52.6% |
| Race - Caucasian | 71.4% | 69.9% |
| Race - African American | 25.1% | 21.8% |
| Ethnicity - Hispanic | 0.3% | 1.2% |
| Brain injury | 2.3% | 1.7% |
| Brain tumor | 0.1% | 0.6% |
| Stroke | 15.9% | 14.1% |

As discussed in the method section, there are multiple objectives for computational phenotyping. First of all, we would like to have a high Area under the ROC curve (AUC) for strong discrimination between groups of interest. Then, we would like to have sparse and less overlapping phenotypes to encourage interpretability. The sparsity was computed as an averaged Gini index of involvement values in each pattern (i.e., the row of *B* and *C*) [21]. The overlap measures the degree of overlapping between all pattern pairs [14]. Lastly, we would also like to see that inferred phenotypes have a small reconstruction error to the original data using mean squared error (MSE). These objectives are controlled by different parameters in our tensor factorization model introduced in the method section, namely $\mu$, $\lambda$, $\gamma$, which corresponds to the weight for discrimination, sparsity, and adherence to co-occurance similarity. After extensive parameter tuning, we found that *R*=30 (number of phenotypes), $\mu = 1$, $\lambda = 0.1$, $\gamma = 1$ performs best in terms of discrimination and interpretability (see Table 4 for some experiements). We implemented regularized coupled TF (Eqn. 2) using Pytorch 11.4 using adaptive momentum estimation (ADAM) for optimization. We set the maximum number of iterations as 1,000. The running time was less than 15 secs with 3 parallel GPUs on DGX-2. We added dropout to logistic regression coefficients for robustness.

**Table 4:** Parameters for different experiments and our criteria of selection. Mean (standard deviation) after 10 trials.

| $\mu$ (logit) | $\lambda$ (l1-norm) | $\gamma$ (similarity) | AUC | Sparsity | Overlap | MSE |
|---|---|---|---|---|---|---|
| 0 | 0 | 0 | 0.586 (0.0138) | 0.2119 (0.0023) | 0.8536 (0.0034) | 0.5148 (0.002) |
| 0 | 0 | 1 | 0.6779 (0.0105) | 0.2442 (0.0038) | 0.8647 (0.0085) | 0.6774 (0.0078) |
| 0 | 0.1 | 0 | 0.5852 (0.0141) | 0.2473 (0.0048) | 0.814 (0.0073) | 0.5215 (0.0027) |
| 1 | 0 | 0 | 0.5999 (0.0106) | 0.2121 (0.004) | 0.8519 (0.0069) | 0.5144 (0.002) |
| **1** | **0.1** | **1** | 0.6925 (0.0182) | 0.2498 (0.0065) | 0.8651 (0.01) | 0.6893 (0.0083) |

We learn a logistic regression model using 30 phenotypes as predictors and select phenotypes that separate epilepsy patients developing AD vs. epilepsy without developing AD with statistical significance (Table 5.a). There are 21 phenotypes showing *p*-value < 0.05 (15 phenotypes for not developing AD and 6 phenotypes for developing AD). To verify the relationship between the phenotypes and epilepsy (EP) progression to AD, we also report the number of patients in the two groups according to the extent of involving to each phenotype (Table 5.b). Due to limited space, we only report three representative phenotypes from each group. We found that in AD likely phenotypes (7, 9, 27) the ratio of AD patients (EP+AD) to EP only patients (EP) increases as the membership values increases. In contrast, in AD unlikely phenotypes (1, 12, 20) the ratio of EP+AD patients to EP only patients decrease as the membership values increases. These findings are consistent with the logistic regression results in which EP+AD patients are more likely to have larger values on AD likely phenotypes and to have smaller values on AD unlikely phenotypes. Note that our cohort is imbalanced with 6,500 EP only patients and 1,289 EP+AD patients, thus the number of EP only patients are always larger than the number of EP+AD across all membership values.

We also make efforts to visualize the phenotypes (Table 6). Each phenotype is represented as a graph that consists of *from*-entities nodes where the transition comes from and *to*-entities nodes where the transition goes to. To present the most highly meaningful edges between the from- and to- nodes, we sort all the edge between the from-entities and to-entities based on *edge scores* and randomly selected top edges with high scores. The edge scores are computed from membership values of entities and transition probability between them, that is, the edge score for $m_1 \rightarrow m_2$ is computed as $B[m_1, r] \times C[m_2, r] \times [log_2(tran(m_1 \rightarrow m_2)) + \varepsilon]$, where $m_1 \in$ *from*-entities, $m_2 \in$ *to*-entities of *r*-th phenotype, and $trans(m_1 \rightarrow m_2)$ is transition probability averaged on all patients; $\varepsilon$ is a positive constant added to the log2-transformed transition probability to make it nonnegative. The edge score will be high if entities are highly involved in the phenotype and transition between the two entities is prevalent. Some of the patterns are very interesting: (1) Potassium chloride is the hub in phenotype 7, which indicates these patients were suffering from, or had a setting for hypokalemia. Hypokalemia may conceivably play a role in epileptic seizures [22,23] and new evidence indicates rubidium and potassium level alterations in Alzheimer's disease [24]. Potassium deficiency can lead to significant decrease in lipid metabolism [25], which strengthens linkage to AD [26].

**Table 5:** Coefficients and *p*-value of phenotypes. Red means AD likely, Green means unlikely to develop AD.

(a) coefficients, *p*-values of phenotypes  (b) Comparison of patient memberships for two types of phenotypes

|     | coeff.  | std. err | z      | P>\|z\| |
|-----|---------|----------|--------|---------|
| x1  | -0.4576 | 0.083    | -5.537 | 0       |
| x2  | 0.1314  | 0.036    | 3.602  | 0       |
| x4  | -0.1465 | 0.069    | -2.127 | 0.033   |
| x5  | -0.3966 | 0.063    | -6.346 | 0       |
| x7  | 0.1587  | 0.045    | 3.553  | 0       |
| x9  | 0.1593  | 0.034    | 4.657  | 0       |
| x11 | -0.1366 | 0.044    | -3.134 | 0.002   |
| x12 | -0.8803 | 0.095    | -9.257 | 0       |
| x15 | -0.1892 | 0.042    | -4.477 | 0       |
| x16 | -0.3983 | 0.083    | -4.773 | 0       |
| x18 | -0.2517 | 0.037    | -6.817 | 0       |
| x19 | -0.3533 | 0.069    | -5.109 | 0       |
| x20 | -0.5505 | 0.077    | -7.175 | 0       |
| x21 | -0.1917 | 0.07     | -2.727 | 0.006   |
| x22 | -0.1519 | 0.028    | -5.42  | 0       |
| x23 | 0.0893  | 0.022    | 3.994  | 0       |
| x24 | 0.0793  | 0.031    | 2.573  | 0.01    |
| x27 | 0.2199  | 0.058    | 3.815  | 0       |
| x28 | -0.4062 | 0.048    | -8.499 | 0       |
| x29 | -0.4162 | 0.085    | -4.906 | 0       |
| x30 | -0.2621 | 0.049    | -5.333 | 0       |

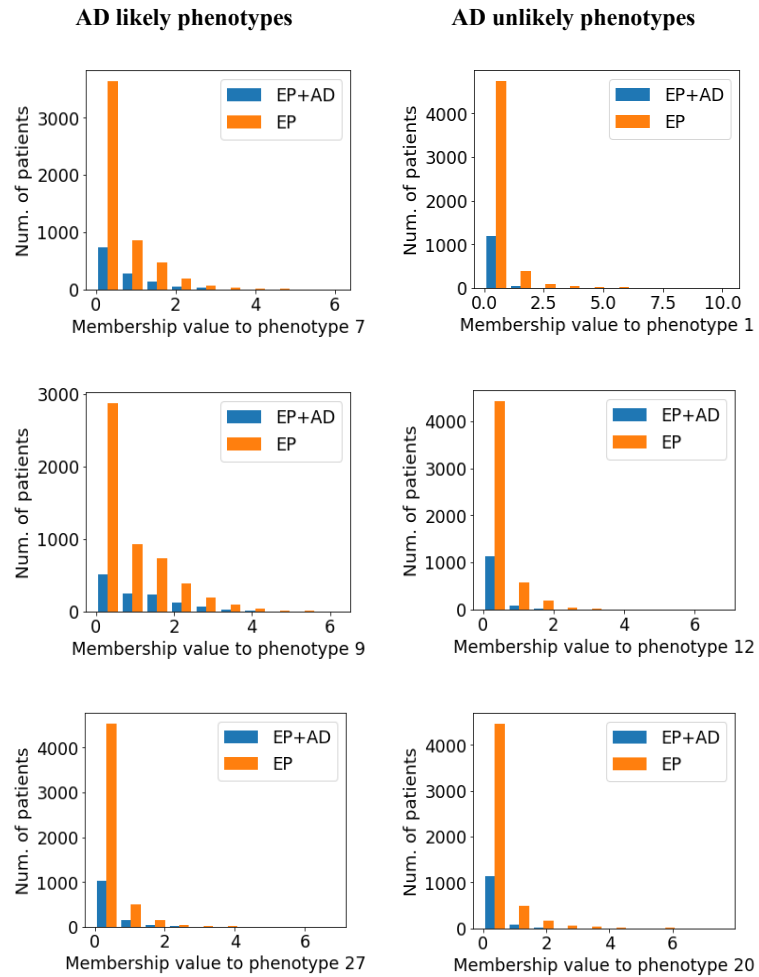

*EP+AD = Epilepsy patients who get AD, EP = Epilepsy patients without AD.

A similar connection from potassium, hypertension to AD seems to exist as an interesting longitudinal study found the use of potassium-sparing diuretics significantly reduced the risk of AD [27]. (2) Phenotype 9 are related to patients who are on Warfarin and disorder of joint, followed by diagnoses on disorder of back and other cerebral degeneration. Many edges in phenotype 9 are similar to those in phenotype 27, but the latter has a deeper structure and patients have disorders of back and additional symptoms involving head and neck. This seems to be indicate the severity of the epilepsy (conditions are likely to be linked with incidental falling), which might be an indicator to the risk of developing AD. (3) Phenotype 23 seems to represent a slightly different group of patients who got diagnosed with epilepsy and recurrent seizures and have immediate problems with urethra and urinary tract, pneumonia, vitamin D deficiency, and mental disorders. We know pneumonia and mental disorders are often diagnosed with late stage AD patients, which may indicate something special about this phenotype group, i.e., patients may have already developed AD (or mild cognitive impairment) but the diagnosis was made later. (4) Phenotype 24 is associated with the medication of piperacillin sodium, which is mainly used to treat pneumonia and skin conditions, followed by diagnosis including other disorders of bones and cartilage, other disorder of soft tissue, etc. We know piperacillin sodium belongs to penicillin antibiotics, and often cause seizures and myoclonus [28]. This may indicate piperacillin aggravation of seizures and increased risk of AD.

**Table 6:** Top transitions in the AD likely phenotypes. Each directed edge means a transition pattern from A to B is observed between two neighboring visits with a high probability within a phenotype. For example, a large proportion of the patients in phenotype 7 has "furosemide" in a visit at time *t* followed by a diagnosis of "other cerebral degeneration" at time *t+1*, which is illustrated by an edge. Box: medication; Oval: diagnosis.

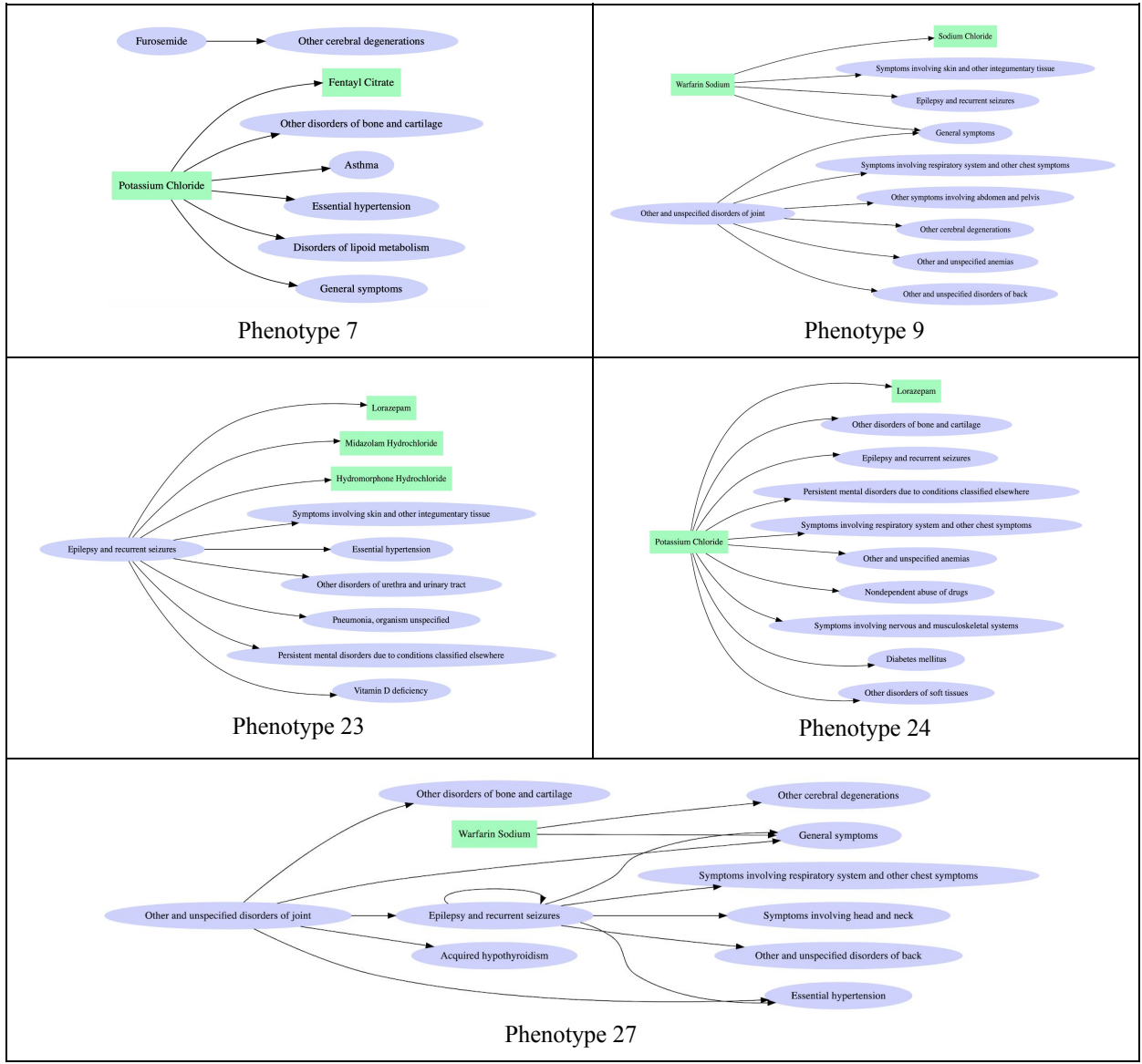

## LIMITATION AND CONCLUSION

Despite the interesting findings and novel contribution of methodology, we admit there are some limitations of this work. We only considered the pairwise transition patterns between consecutive visits, but patients might visit hospitals outside of our network and therefore causing bias to our model. Certain long term effect can play an important role in the pathway from epilepsy to Alzheimer's disease might be neglected in the 3d tensor factorization. It is possible to integrate high-order terms in tensor factorization but it is not efficient to consider every combination because most high order transition patterns are invalid. This will be a future extension of our. In summary, we confirmed the strong causal relationship between epilepsy and Alzheimer's disease using big data, and conducted integrative computational phenotyping analysis (AUC~0.7) to discriminate AD likely and unlikely phenotypes among epilepsy patients, providing signature transition patterns that might inspire clinical research. A detailed analysis of the exact phenotypic interactions with epilepsy and subsequent AD is necessary. In this study, we only analyzed epilepsy subgroups leading to AD but not AD subgroups leading to epilepsy, which can be another direction to explore in the future.